\documentclass[aps,twocolumn,pre,showkeys,showpacs,eqsecnum,citeautoscript]{revtex4-1}
\usepackage{graphpap}
\usepackage[dvips]{graphicx}
\usepackage[dvips]{graphics}
\usepackage{color}
\usepackage{amsmath, amssymb, amsfonts, amsthm}
\usepackage{bbold}
\usepackage[pagebackref=false]{hyperref}

\begin{document}

\title{Loading a quantum-dot based ``Qubyte'' register}

\author{C. Volk$^{1*}$, A.M.J. Zwerver$^{1*}$, U. Mukhopadhyay$^{1}$, P.T. Eendebak$^{2}$, C.J. van Diepen$^{1}$,\\ J.P. Dehollain$^{1}$, T. Hensgens$^{1}$, T. Fujita$^{1x}$, C. Reichl$^{3}$, W. Wegscheider$^{3}$ and L.M.K. Vandersypen$^{1\diamond}$}
\affiliation{$^1$QuTech and Kavli Institute of Nanoscience, TU Delft, 2600 GA Delft, The Netherlands\\
$^2$QuTech and Netherlands Organization for Applied Scientific Research (TNO), 2600 AD Delft, The Netherlands\\
$^3$Solid State Physics Laboratory, ETH Z\"{u}rich, 8093 Z\"{u}rich, Switzerland\\
$^*$ These authors contributed equally to this work\\
$^x$ Present address: The Institute of Scientific and Industrial Research, Osaka University, Ibaraki, Osaka, Japan\\
$^\diamond$ Corresponding author. Email: l.m.k.vandersypen@tudelft.nl}

\date{ \today}

\begin{abstract}
Electrostatically defined quantum dot arrays offer a compelling platform for quantum computation and simulation. However, tuning up such arrays with existing techniques becomes impractical when going beyond a handful of quantum dots. Here, we present a method for systematically adding quantum dots to an array one dot at a time, in such a way that the number of electrons on previously formed dots is unaffected. The method allows individual control of the number of electrons on each of the dots, as well as of the interdot tunnel rates. We use this technique to tune up a linear array of eight GaAs quantum dots such that they are occupied by one electron each. This new method overcomes a critical bottleneck in scaling up quantum-dot based qubit registers.
\end{abstract}

\maketitle

\newpage
\section{Introduction}

Quantum-dot based electron spin qubit systems~\cite{Loss1998,Hanson2007, Vandersypen2017} have made significant steps towards becoming a scalable platform for quantum computation. Important landmarks include the realization of 99.9\%-fidelity single-qubit gates~\cite{Yoneda2018}, the implementation of two-qubit gates~\cite{Nowack2011, Shulman2012, Veldhorst2015, Zajac2018, Huang2018, Xue2018} and two-qubit algorithms~\cite{Watson2018}.
Although a high degree of control of the charge and spin degrees of freedom has been shown, research has been mainly limited to single, double and triple dot systems. Recently, control of the charge occupation of four dot systems has been demonstrated ~\cite{Thalineau2012, Takakura2014, Fujita2017, Mukhopadhyay2018} and a single electron could be controllably placed in a 3x3 array~\cite{Mortemousque2018}. However, device specific approaches to tuning quantum dots will need to be replaced by  a systematic approach, as arrays become larger with the scale-up of quantum-dot based quantum circuits.

The controlled formation and filling of large quantum dot (QD) arrays poses multiple challenges. Individual gate voltages affect not only the parameter they are designed to control, typically the electrochemical potential of a specific QD or the tunnel barrier between two adjacent QDs, but through capacitive cross-talk also affect other electrochemical potentials and tunnel barriers~\cite{vanderWiel2002}. Furthermore, tuning  devices is complicated by a disordered potential landscape arising from charges trapped in randomly located impurities and defects in the substrate and at the surface~\cite{Pioro2005,Buizert2008}. Finally, electrons are loaded into QDs from an electron reservoir. When a target dot is separated from the reservoir by one or more other dots, electrons are typically loaded by co-tunneling, only virtually occupying the intermediate dots. However, for more than three or four dots, the co-tunnel rates become impractically low.

These challenges present themselves when measuring the charge occupation in quantum dot arrays through conventional charge stability diagrams. In such diagrams, the signal from a charge sensor is recorded while sweeping two gate voltages, resulting in a 2D plot that exhibits regions in gate voltage space with a fixed number of electrons on each dot, separated by lines indicating charge additions to the array, or charge transitions between dots~\cite{vanderWiel2002}. Such a 2D plot corresponds to a plane in a multi-dimensional space spanned by all the gate voltages. As arrays get larger, when sweeping just two gate voltages, cross-talk leads to slopes of charge transition lines that are almost parallel and hard to distinguish. Assignment of charge transition lines to specific dots is further complicated by non-uniform addition energies. Furthermore, the intersections between different charge addition lines can cluster together in a small gate voltage region. Finally, the difficulty of loading electrons to dots far from the reservoir leads to postponed loading of dots (latching) or to missing charge addition lines~\cite{Yang2014}. Those complications lead to plots that are difficult to interpret~\cite{Ito2016}.

Cross-talk and the background disorder potential have been compensated for in short dot arrays using so called virtual gates, which are linear combinations of multiple gate voltages chosen such that only a single electrochemical potential or tunnel barrier is addressed~\cite{Hensgens2017}. Virtual gates also make it possible to strategically choose the measured 2D plane in gate-space, so that multi-dot charge stability diagrams become easy to interpret~\cite{Hensgens2017,Mortemousque2018,Medford2013}.
The difficulty of loading electrons into large arrays has been circumvented using additional reservoirs in between groups of three dots~\cite{Malinowski2018}. In another approach, an additional access point to a reservoir was created halfway a linear array of five QDs~\cite{Ito2016}. Instead of loading electrons by co-tunneling, electrons can also be made to sequentially tunnel through a chain of dots to reach their target location~\cite{Baart2016a}, but this approach requires the chain of dots to be already formed in the first place.

We explored several approaches to form long linear arrays in a controlled way, such as forming individual single dots first and stitching them together, stitching together double dots, or starting with a large QD and then splitting it up into an array of separate dots. However, we found it difficult to make these approaches work well.

Here, we show the controlled filling of an array of eight QDs, which we call a Qubyte register, using a  method that is both conceptually simple and effective. Starting from a double dot, we introduce the ``$n+1$ method'', adding dots one by one using virtual gates. Every new dot added adjacent to the existing array is right next to a reservoir so the dot can be filled easily. The use of virtual gates saves the charge occupation in the previously formed dots while adding a new dot, and also keeps the charge stability diagrams simple to interpret. We show that we can locally control the number of electrons on each dot down to the last electron, and that we can set all interdot tunnel couplings to typical values used in spin qubit experiments. Finally, we discuss the limitations and potential pitfalls of the $n+1$ method.

\section{Results}
\subsection{Device and Initial Characterization}

\begin{figure*}
\centering
\includegraphics [width=0.9\linewidth] {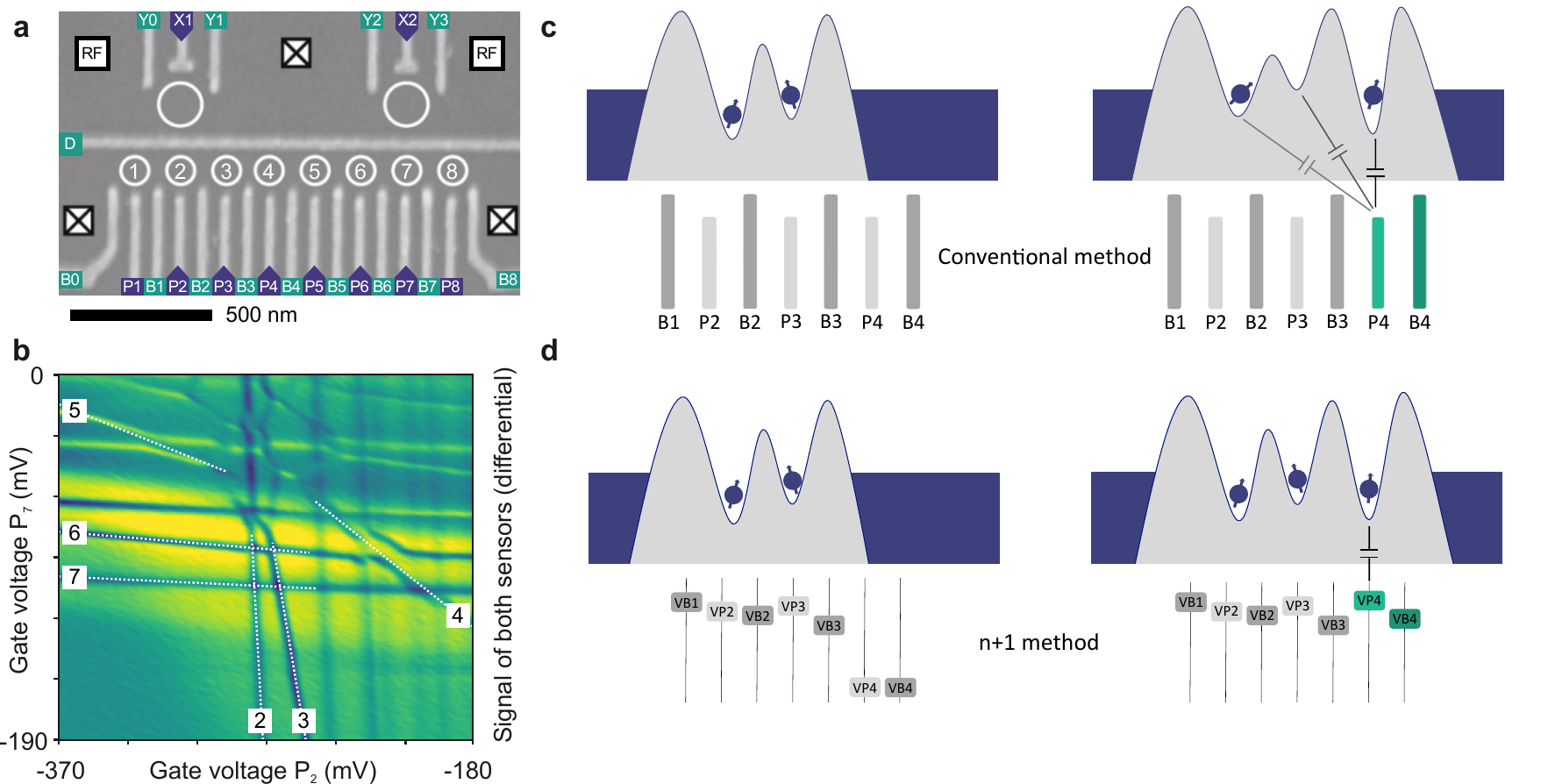}
\caption[]{\textbf{Device design and tuning principle.}
(a) Scanning electron micrograph of a device nominally identical to the one used in the experiment. The circles indicate the intended positions of eight quantum dots (QDs) that define a Qubyte register and of  two additional dots that are used for charge sensing. For the linear array, the designed dot-to-dot pitch is 160~nm. The plunger gates connected to high frequency lines are marked with blue triangles. The squares indicate the position of the Fermi reservoirs. Two on-board tank circuits for RF reflectometry readout are connected to each of the sensing dots.
(b) Charge stability diagram of a sextuple dot formed between barrier gates $B_1$ and $B_7$. The sum of the differential demodulated voltages of both sensing dots is plotted. The dashed lines highlight charge transitions of each of the six QDs (the numbers refer to the labels in panel (a)).
(c) Illustration of the potential landscape of a double QD. Gates $P_4$ and $B_4$ are used to form a third QD. Due to capacitive cross-talk, indicated by the capacitor symbols, these gates influence the potential of the other QDs as well (to avoid clutter, we did not draw any other capacitor symbols).
(d) A double QD is extended to a triple QD using the virtual plunger $\mathrm{\mathrm{VP}}_4$ and barrier $\mathrm{VB}_4$. Due to cross-capacitance compensation these parameters only act locally on the potential
landscape.
}
\label{fig1}
\end{figure*}

Fig.~\ref{fig1}(a) shows a scanning electron micrograph of a device nominally identical to the one used in the experiment. The gate layout has been adapted from previous triple and quadruple quantum dot
devices~\cite{Medford2013,Fujita2017}.
On one side, 17 gates with a pitch of 80~nm are fabricated to control the tunnel barriers and electrochemical potentials of the QDs. The upper part of the sample accommodates two sensing dots (SD) that are capacitively coupled to the linear QD array. The circles indicate the intended positions of the QDs. All measurements are carried out in a dilution refrigerator with a base temperature below 20 mK.

Initially, the device is characterized by DC transport measurements.
The pinch-off characteristics of the channel between each of the plunger $P_i$ or barrier $B_i$ gates and the central gate $D$ is measured (see schematics in Fig.~\ref{fig1}(a)) and single QDs are formed by sweeping pairs of neighbouring barrier gates. These measurements confirm that all QDs, including the sensing dots, can be formed. Moreover, the pinch-off values determined for each gate act as starting parameters for further tuning.
In all subsequent measurements, we probe the linear QD array via the two sensing dots, which are sensitive to the number of electrons in the array, as well as to their position in the array. The charge sensors are probed using RF reflectometry (see \hyperref[Methods]{Methods} section).

To illustrate the difficulty of traditional tuning strategies, Fig.~\ref{fig1}(b) shows a charge stability diagram for a linear six-dot array (sextuple dot) confined between the barrier gates $B_1$ to $B_7$. The charge stability diagram has been recorded sweeping the voltages of gates $P_2$ and $P_7$, i.e.\ the gates mostly coupled to the outer QDs. In the diagram charge addition lines with different slopes can be identified. However, charge transitions with similar slopes can be only be assigned unambiguously to specific dots, after also stepping other gate voltages (see e.g. the small difference in slope between the transitions for dots 6 and 7).
Even then, the complex pattern of transitions in the center of the diagram makes it extremely difficult to determine the charge occupation at every point in this gate space. Moreover, cross-capacitances hinder local tuning of the electrochemical potential and tunnel rates.

\subsection{$n+1$ method}
To tune up a multi-dot array dot by dot, we make use of virtual gates, which compensate for the cross-talk that occurs when sweeping actual gate voltages (see Fig.~\ref{fig1}(c-d)). The virtual gates as used here
compensate for cross-talk on the electrochemical potentials only, not for cross-talk effects on tunnel barriers.
The virtual plunger gate VP$_i$ directly corresponds to the electrochemical potential of QD$_i$, up to a lever arm. The compensation is performed to first order, so that we can express the virtual gates as linear combinations of the physical gate voltages, summarized by a cross-capacitance matrix~\cite{Nowack2011,Hensgens2017}.

The tuning procedure consists of the following steps, described in more detail below:\\
1.\@ Tune up a double quantum dot (DQD) and the two sensing dots with the traditional strategy (Fig 2(a)).\\
2.\@ Measure the cross-capacitance between all gates and the electrochemical potentials of these four QDs and record them in a cross-capacitance matrix. This matrix can now be used to generate virtual gates (Fig 2b).\\
3.\@ Use the virtual plunger and barrier gates adjacent to the existing dots to form the next QD without disturbing the former.\\
4.\@ Measure the cross-capacitance of every gate to the newly formed QD and place these values in the corresponding row of the matrix.\\
5.\@ Re-measure the cross-capacitances to the previously formed QDs and update the matrix accordingly.\\
Steps 3 to 5 are repeated to extend the array, adding one QD at a time.\\

\begin{figure}
\centering
\includegraphics [width=1\linewidth] {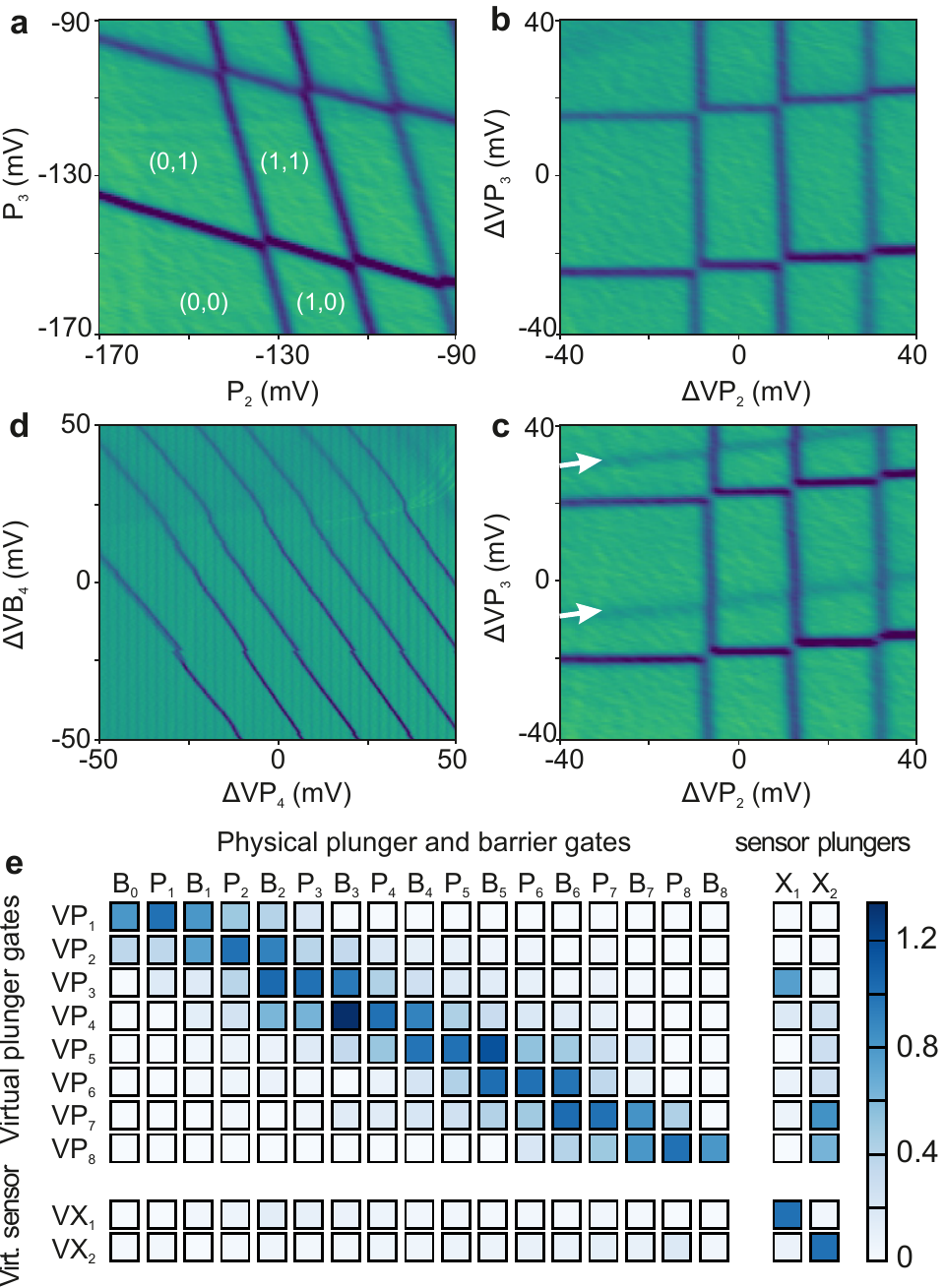}
\caption[]{\textbf{Tuning method.}
(a) Charge stability diagram of a DQD in the single electron regime. The charge sensor response is plotted in color scale (the differential demodulated voltage, in arbitrary units, is plotted here and in similar plots below) as a function of the plunger gate voltages $P_2$ and $P_3$.
(b) Charge stability diagram of the same DQD recorded as a function of the virtual plungers $\mathrm{VP}_2$ and $\mathrm{VP}_3$.
(c) Charge stability diagram where an additional QD has been formed to the right of the DQD by raising the relevant tunnel barrier. The arrows indicate the position of the addition lines of the newly formed dot.
(d) Differentiated demodulated charge sensor signal as a function of virtual plunger $\mathrm{VP}_4$ and virtual barrier $\mathrm{VB}_4$. The charge addition lines corresponding to the newly formed QD are clearly visible. No transitions of the pre-existing dots are observed, due to the use of virtual gates.
(e) Visualization of the cross-capacitance matrix of the eight-dot array. The entries of each row show how the virtual plunger value (and hence the electrochemical potential) of a QD is influenced by other gate voltages. The rows for virtual barrier gates are omitted for simplicity. The plungers of both sensing dots are included in the matrix.
}
\label{fig2}
\end{figure}

The matrix entries for the initial DQD ($\rm{QD}_i$ and $\rm{QD}_{i+1}$) are determined by how much an addition line for $\rm{QD}_i$ in a $\rm{P}_i$ scan is displaced when stepping any of the other plunger (barrier) gates $P_j$ ($B_j$) by an amount $\delta V$ (see Supplementary Fig.~\ref{figS1}). The ratio of the shift of the charge transition line of $\rm{QD}_i$ in the $\rm{P}_i$ scan and $\delta V$ yields the corresponding entry in the cross-capacitance matrix. We do this for all eight plunger and nine barrier gates of the linear array, as well as for the plunger gates of the sensing dots.

We illustrate how the cross-capacitance matrix is used to create virtual gates for the first three dots (leaving out the outer barrier gates for simplicity). The following relationship expresses how much each physical gate affects each virtual gate:
\begin{equation*}\label{matrix}
  \left(\begin{array}{c} \mathrm{\Delta VP}_1 \\ \mathrm{\Delta VB}_1 \\ \mathrm{\Delta VP}_2 \\ \mathrm{\Delta VB}_2 \\ \mathrm{\Delta VP}_3 \end{array}\right) = \left(\begin{array}{ccccc} 1 & \alpha_{12} & \alpha_{13}   &\alpha_{14} &\alpha_{15}\\ 0 & 1 & 0 & 0 & 0 \\ \alpha_{31} & \alpha_{32} & 1 &\alpha_{34} &\alpha_{35} \\ 0 & 0 & 0 & 1 & 0 \\ \alpha_{51} & \alpha_{52} & \alpha_{53} &\alpha_{54} & 1 \end{array}\right)   \left(\begin{array}{c} \Delta P_1 \\ \Delta B_1 \\ \Delta P_2 \\ \Delta B_2 \\ \Delta P_3 \end{array}\right)
\end{equation*}
For convenience, we set the diagonal entries to 1 (dimensionless), disregarding the lever arm. This implies we express virtual gates in units of Volt, similar to the physical gates. Furthermore, since we do not include cross-talk effects on tunnel barriers, the off-diagonal matrix elements relating the physical gate voltages to virtual barrier gates are set to zero. The inverse matrix expresses the linear combination of physical gate voltages that is needed to sweep a virtual gate. We note that the diagonal entries of the inverse matrix do not need to be equal to 1.

The effectiveness of the cross-talk compensation can be seen by recording a charge stability diagram in the virtual gate space, i.e. using $\mathrm{VP}_i$ and $\mathrm{VP}_{i+1}$ as sweep parameters (see Fig.~\ref{fig2}(b)). Ideally, addition lines of $\rm{QD}_{i}$ and $\rm{QD}_{i+1}$ appear as orthogonal (horizontal and vertical) lines. In practice, the compensation is not always perfect because we extrapolate each cross-capacitance from just two data points (see Supplementary Fig.~\ref{figS1}), but it is usually good enough.

To extend the DQD to a triple dot, we form a new tunnel barrier using a neighbouring virtual barrier gate, e.g. $\rm{VB}_{i+2}$. The pinch-off values determined in DC transport indicate a suitable voltage range to scan with the barrier gate. Optionally, we then monitor the charge stability diagram $\mathrm{VP}_i$ - $\mathrm{VP}_{i+1}$ while stepping $\rm{VB}_{i+2}$. Once the barrier is raised sufficiently to form an additional QD, new addition lines appear in the charge stability diagram (see arrows in Fig.~\ref{fig2}(c)). The charge transitions of the previously tuned QDs are only slightly affected, indicating the effectiveness of the virtual gate concept.

We complete the tuning of the newly formed dot to the single electron regime by measuring a charge stability diagram sweeping  virtual plunger $\mathrm{VP}_{i+2}$ versus  virtual barrier $\mathrm{VB}_{i+2}$. A set of diagonal lines indicates charge transitions of the newly formed QD (see Fig.~\ref{fig2}(d)). We can identify the last charge transition in the bottom left of the figure.

The next step is to update the cross-capacitance matrix. First, we fill the row corresponding to VP$_{i+2}$. The effect of all $\mathrm{VP}_{j}$ and $\mathrm{VB}_{j}$ on $\mathrm{VP}_{i+2}$ is determined, as described for the first double dot, with the distinction that we now start from a set of virtual gates, expressed by matrix $A$.

As a final step, the existing matrix entries are updated to account for reduced screening of the gate potentials when the two-dimensional electron gas at the location of QD$_{i+2}$ is depleted. To do so, we remeasure the cross-talk from all the virtual plunger and barrier gates to all the virtual plunger gates. This results in a matrix $A^{\prime}$. The updated cross-capacitance matrix is found by matrix multiplication $A_{new} = A^\prime A$.

\subsection{Qubyte in the single electron regime}

\begin{figure*}
\centering
\includegraphics [width=0.9\linewidth] {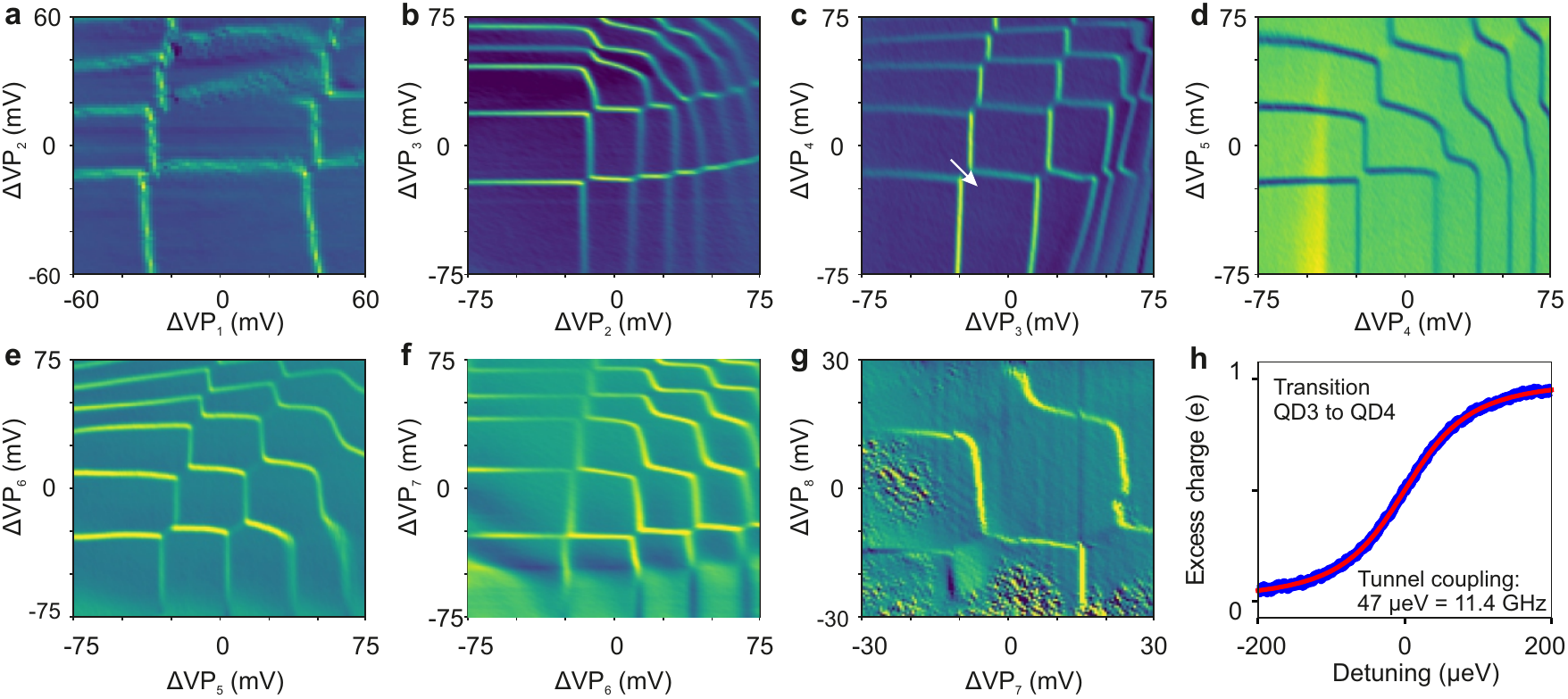}
\caption[]{\textbf{Qubyte in the single electron regime.}
(a-g) Charge stability diagrams of pairs of neighbouring QDs within the array. The differential demodulated voltage is plotted as a function of the virtual plunger gates. For panels (b-f), the charge stability diagrams were measured with a sextuple QD  defined between barrier gates $B_1$ and $B_7$. The measurements are centered around the single electron occupation.
Panels (a,g) show charge stability diagrams of the two outermost DQDs of an octuple QD array.
(h) Scan along the interdot detuning axis of $\mathrm{QD}_3$ and $\mathrm{QD}_4$ with one electron in those two dots (see arrow in (c)). The data has been fit according to the model described
in~\cite{DiCarlo2004,Hensgens2017}.
}
\label{fig3}
\end{figure*}

We apply the $n+1$ method to form a sextuple QD and octuple QD. We start with a DQD confined between the barrier gates $B_1$ and $B_3$ and initially extend the array to the right. The electrochemical potentials and thus the number of electrons residing on all QDs can be independently controlled. The results are verified by charge stability diagrams of neighbouring pairs of virtual plunger gates, see Fig.~\ref{fig3}(b-f), where the sextuple dot has been initialized with one electron in each of the QDs. The gate voltages at the center of all of these plots are identical. All data sets have been acquired by fast voltage sweeps. At low resolution and low averaging, sufficient for initial tuning, the acquisition time per panel is on the order of a few 100~ms. High-quality data such as those shown in Fig.~\ref{fig3}(b-f) take approximately 10~s per panel.
Each plot can be interpreted as a charge stability diagram of a DQD, independent of the neighbouring QDs. The virtual gates control the electrochemical potential of the DQD and the number of electrons can be determined easily from the measurements. This set of measurements contains the full information of the charge state of the sextuple QD and is much easier to interpret and work with than conventional charge stability diagrams, where multiple charge addition lines as well as interdot transitions are visible in a single plot. In our experience, this new method renders the conventional charge stability diagram obsolete. In fact, the data of Fig.~\ref{fig1}(b) was taken for illustration purposes only, after forming the sextuple dot using measurements such as those in Fig.~\ref{fig3}.

Following the same $n+1$ method, the sextuple QD is further extended to an octuple QD array by adding another QD on each side. Due to limitations of the experimental setup, the plunger gates $P_1$ and $P_8$ are not connected to high-frequency lines necessary to apply fast gate voltage sweeps. Therefore, any measurement involving these gates must rely on slow gate voltages sweeps, in practice with a cut-off frequency below 1~Hz. For this reason, we first formed a sextuple dot in the center and only then extended it to an octuple dot. Fig.~\ref{fig3}(a) and (g) show charge stability diagrams as a function of $\mathrm{VP}_1$, $\mathrm{VP}_2$ and $\mathrm{VP}_7$, $\mathrm{VP}_8$, respectively, completing the formation of the Qubyte register.

The cross-capacitance matrix for the octuple QD configuration of Fig.~\ref{fig3} is shown in Fig.~\ref{fig2}(e). It visualizes the effect of plunger and barrier gates on the electrochemical potential of all QDs. As discussed, each row has been normalized such that the diagonal elements are 1. In these units, the effect of the closest barrier gates on the electrochemical potential of a QD is typically between 0.9 and 1.1. This is in agreement with the device geometry (see Fig.~\ref{fig1}(a)) where the barrier gates are 30~nm longer than the plunger gates, bringing them close to the expected QD position. The influence of a neighbouring plunger gate on a dot potential is on the order of 0.4 to 0.5 and the one of the next-nearest neighbour 0.15 to 0.2, so the coupling diminishes with distance, as expected (Supplementary Fig.~\ref{figS2} plots the cross-capacitance versus distance).
The cross-capacitance to the sensing dots is small (typically below 0.1), but nevertheless it is relevant to correct for, as the sensing dots are operated at a steep slope of a Coulomb peak to maximize the charge detection sensitivity.

By means of the virtual barrier gates $\mathrm{VB}_i$, we can adjust the interdot tunnel couplings while cross-capacitance correction compensates the influence on the electrochemical potentials. To determine the interdot tunnel coupling, we measure the charge sensor response along the detuning axis across a single-electron transition. Fig.~\ref{fig3}(h) shows an example for the (1,1,0,1,1,1)-(1,0,1,1,1,1) transition, where the numbers in brackets indicate the number of charges on each of the six dots, from QD$_2$ to QD$_7$. The data is fit according to a simple model considering broadening of the transition due to tunnel coupling and thermal excitation~\cite{DiCarlo2004,Hensgens2017}, using a measured effective electron temperature of $T_e =$~90~mK. The tunnel coupling for all pairs of neighbouring dots has been tuned to a range of 5 to 15~GHz (see Supplementary Fig.~\ref{figS3}).

\begin{figure}
\centering
\includegraphics [width=1\linewidth] {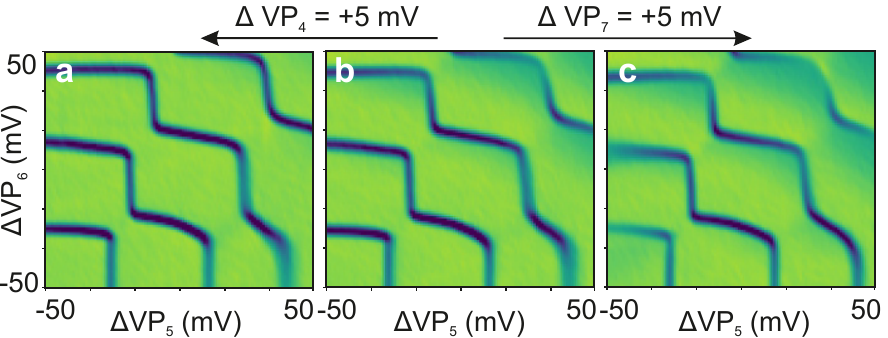}
\caption[]{\textbf{Verifying independent control of the electrochemical potentials.}  Charge stability diagrams of $\rm{QD}_5$ and $\rm{QD}_6$ measured as function of the virtual plunger gates with all other QDs near the center of a charge stability region. Panel (b) is a reference measurement. Panel (a) and (c) are taken for identical gate voltages as (b) except that  $\rm{VP}_4$ ($\rm{VP}_7$) is increased by 5 mV. We note that stepping $\rm{VP}_4$ ($\rm{VP}_7$) has almost no influence on the charge stability diagram of $\rm{QD}_5$ and $\rm{QD}_6$. This figure shows a subset of the data shown in Supplementary Fig.~\ref{figS4}.
}
\label{fig4}
\end{figure}

To further verify the validity of the $n+1$ method implemented via the use of virtual gates, we record the charge stability diagram of two neighbouring dots, while all other dots are kept in Coulomb blockade. One by one we step the virtual plunger gates of the neighbouring dots, which ideally should not affect the measured charge stability diagram. Fig.~\ref{fig4} depicts such a test for $\rm{QD}_5$ and $\rm{QD}_6$. In panel (a) $\rm{VP}_4$ has been increased compared to panel (b) and in panel (c) $\rm{VP}_7$ has been increased. The charge stability diagram is not affected by small changes in neighbouring electrochemical potentials, which implies that the virtual gates behave as expected and verifies that the charge stability diagram indeed shows addition lines of the expected dots. The same measurements are repeated for all QDs; charge stability diagrams of neighbouring QDs were measured while the electrochemical potential of all other QDs has been altered. Data sets for all gate combinations are shown in Supplementary Fig.~\ref{figS4}, showing similar results as presented in Fig.~\ref{fig4}.

\begin{figure}
\centering
\includegraphics [width=1\linewidth] {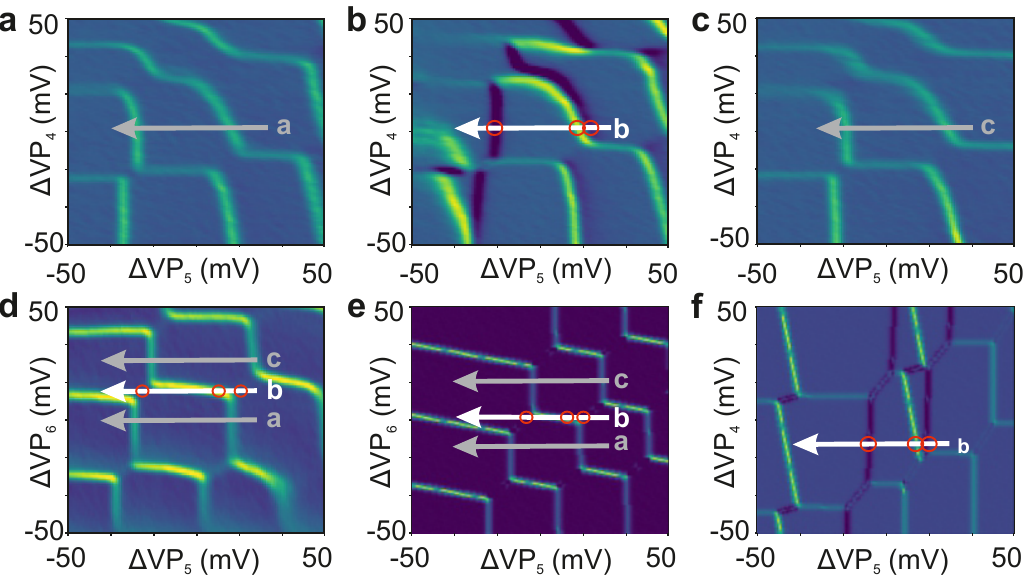}
\caption[]{\textbf{Possible pitfalls when using virtual gates.}
(a-c) Charge stability diagrams of $\rm{QD}_5$ and $\rm{QD}_4$. The virtual plunger $\mathrm{VP}_6$ is stepped by 10~mV from (a) to (b) and by another 15~mV from (b) to (c). In (b), additional lines appear, which arise both from interdot and dot-reservoir transitions, as explained in the text.
(d) Charge stability diagram of $\rm{QD}_6$ and $\rm{QD}_5$. The arrows indicating cuts along the $\mathrm{VP}_5$ axis for three different values of $\mathrm{VP}_6$ correspond to the arrows in panels (a-c).
(e) Simulated double dot charge stability diagram computed for a triple quantum dot system. The simulation assumes the constant interaction model, taking into account the capacitive interdot coupling and using on purpose an imperfect cross-capacitance matrix. The simulation results exhibit similar features as the measurements in panel (d). The three arrows are drawn at the corresponding locations as in panel (d) as well.
(f) Simulated charge stability diagram computed for the same triple quantum as in (e), but for the other DQD. Similar to the data in panel (b), we observe several additional interdot and dot-reservoir charge transition lines.
}
\label{fig5}
\end{figure}

We note that it is not trivial that this method works flawlessly and care has to be taken to ensure the electron occupation of each dot is as intended. Specifically, it is important that the neighbouring QDs remain sufficiently far from any charge transitions. This requires that the cross-capacitances are measured with a reasonable accuracy, and that the neighbouring QDs be detuned from the Fermi level by more than the interdot capacitive coupling energy. To illustrate this point, a set of charge stability diagrams for $\rm{QD}_4$ and $\rm{QD}_5$ is shown in Fig.~\ref{fig5}(a-c), with increasing values for $\mathrm{VP}_6$ per panel (A video available as supplementary information shows a similar series of charge stability diagrams in steps of 0.5 mV in VP$_6$.). Fig.~\ref{fig5}(a) shows a reference plot of a clean charge stability diagram. In Fig.~\ref{fig5}(b), the same gate voltages are scanned but VP$_6$ has been changed by 10 mV. Extra lines appear, which disappear again when increasing VP$_6$ further (Fig.~\ref{fig5}(c)). The extra lines can be understood if we inspect the charge stability diagram for $\rm{QD}_5$ and $\rm{QD}_6$, which is depicted in Fig.~\ref{fig5}(d) with arrows indicating the values of $\rm{QD}_6$ used in panels (a-c). We see that arrow b, which corresponds to the case of Fig.~\ref{fig5}(b), passes through an interdot transition of $\rm{QD}_5$ and $\rm{QD}_6$, then intersects an addition line for QD$_6$ (since the virtual gates are not perfect, this addition line is slightly tilted) and finally cuts through another interdot transition of QD$_5$ and QD$_6$. These three crossings occur at the positions of the red circles in Fig.~\ref{fig5}(b). By comparison, arrows a and c do not pass through any charge transitions involving QD$_6$. This set of data makes clear how to avoid ambiguity in controlling and verifying the number of electrons on each dot.

We can observe the same effects in classical simulations of the charge stability diagrams. The simulation considers only three QDs and adopts the constant interaction model~\cite{vanderWiel2002}, meaning the charging energies and capacitive interdot coupling energies are assumed to be constant. Imperfections of the cross-capacitance matrix are taken into account in the model. Other effects, e.g. tunnel coupling, non-linearities of the cross-talk and latching effects are neglected. Fig.~\ref{fig5}(e) shows a simulated charge stability diagram for QD$_5$ and QD$_6$, with the arrows a, b and c at similar locations as in the measurements of Fig.~\ref{fig5}(d). Fig.~\ref{fig5}(f) shows the simulated charge stability diagram for QD$_4$ and QD$_5$, for the case of arrow b. Similar to the data in Fig.~\ref{fig5}(b), we observe extra lines in the simulated charge stability diagram, as arrow b passes through interdot transitions and an addition line for QD$_6$. While details vary, in part because tunnel coupling is not included in the simulation, the simulation results are in good qualitative agreement with the experimental data.

\section{Discussion}
We developed a powerful technique to tune an array of QDs one by one and load it in the few electron regime. We apply this method to tune up a linear array of eight quantum dots in GaAs from scratch. This currently takes one to two days for an experienced user and a well behaved sample. Additionally, the virtual gates technique facilitates tuning of the tunnel couplings, which we showed could be tuned to a relevant range for qubit operations in this device.

With regards to the scalability of this method, we make the following observations. First, the cross-capacitance quickly drops with distance between the gates and the dots. Therefore, only the entries near the diagonal of the  cross-capacitance matrix are relevant and need to be determined. This implies a linear scaling of the number of cross-capacitance elements as a function of the number of dots. Second, as we relied on charge addition lines in charge stability diagrams of neighbouring QDs to determine the number of electrons per dot, each of the QDs must be able to exchange electrons with at least one of the reservoirs. QDs not positioned at the end of the array need to exchange electrons via co-tunneling, mediated by a virtual occupation of the QDs in between~\cite{vanderWiel2002}. The co-tunnel rate scales inversely proportional with the number of interdot tunnel barriers between a dot and the nearest reservoir, as well as with the detuning of the dots in between them~\cite{Braakman2013}. However, this is by no means a fundamental obstacle. When dots are formed one at a time, the newly formed dot is immediately adjacent to a reservoir and can thus be easily loaded. For dots in the interior, the $n+1$ method we introduced in principle takes care of maintaining their occupation through cross-talk compensation. If desired, verifying the dot occupation in the interior of a long array after it is formed can still be done, for instance by emptying the array (while not removing it), followed by sequential tunneling of electrons to the desired locations~\cite{Baart2016a}. Finally, we believe that the $n+1$ method is not bound to a specific device geometry or material. In particular, we expect that it is directly applicable to linear arrays in silicon based QD devices~\footnote{In the course of preparing this manuscript, we became aware of closely related work on a linear array of nine Si/SiGe QDs, see~\cite{Mills2018}} and can be extended to two-dimensional QD arrays. The $n+1$ method can become a standard method to conveniently tune QD arrays and should lend itself well to automation~\cite{Baart2016c,Botzem2018}.

The data also shows the limitations of the current approach. We correct for the cross-capacitance of plunger and barrier gates influencing electrochemical potentials but not for the influence on tunnel barriers. As a consequence, altering a virtual plunger gate will affect neighbouring barriers, as can be seen in Fig.~\ref{fig3}(b). Increasing $\mathrm{VP}_2$ and $\mathrm{VP}_3$ increases the interdot tunnel coupling, which can be deduced from the broadening of the interdot transitions. In principle cross-capacitance effects on barriers can also be taken into account, as was demonstrated recently for a triple dot array~\cite{Hensgens2017}. However, this task is not trivial since the dependence of gate voltage to tunnel coupling is typically exponential and thus the linear approximation of the cross-capacitance matrix is only valid over a limited voltage range. As we have shown in this work, adjusting the interdot tunnel couplings individually is not a very difficult task, and this can be implemented using automated tuning algorithms as well~\cite{VanDiepen2018}.

Altogether, the $n+1$ method shown here enables future experiments involving increasing numbers of electron spin qubits in semiconductor quantum dot arrays. It addresses an important bottleneck in scaling up quantum dot arrays and highlights the potential of this approach for large-scale quantum computation and simulation.

\section{Methods}\label{Methods}
The sample is fabricated from a silicon-doped GaAs/AlGaAs quantum well grown by molecular beam epitaxy. A two-dimensional electron gas is formed 90 nm below the surface. It shows a mobility of $1.6\cdot10^6$cm$^2$/Vs at an electron density of 1.9x10$^{11}$ cm$^{-2}$. A single layer of metallic gates (Ti/Au), defined by electron-beam lithography, is biased with appropriate voltages to selectively deplete the 2DEG underneath.
During cooldown the gates have been biased individually with positive voltages between +50 and +250 mV to reduce charge noise \cite{Pioro2005} and to improve the uniformity of the pinch-off characteristics of the gates. Details are shown in The Supplementary Table~I.

Gates $P_2$ to $P_7$ of the linear array and the plunger gates of both sensing dots ($X_1$ and $X_2$) are, via bias-tees on the printed circuit board (see Fig.~\ref{fig1}(a)), connected to high-frequency lines, which allows combining DC voltages and nanosecond gate voltage pulses on the same gate. The other gates are connected to DC lines.

Except for the initial characterization using DC transport, RF reflectometry is used, enabling fast, simultaneous read out of both charge sensors by frequency multiplexing~\cite{Barthel2010, Hornibrook2014}. As the capacitive coupling and thus the sensitivity decreases with distance from the sensor, we read out both sensors simultaneously to maximize the readout quality. The charge stability diagrams shown in Figs.~\ref{fig2}-\ref{fig5} show the signal from the nearest charge detector. The sum of the derivative along both axis is plotted. In Fig.~\ref{fig1}(b), the signals from the two charge sensors are added. LC tank circuits based on home-built superconducting NbTiN inductors are connected to the ohmic contacts of the sensing dots (see labels RF in Fig.~\ref{fig1}(a)). RF tones close to the resonance frequencies of the tank circuits, at 108.5 MHz and 171.9 MHz, are sent to the sample. The reflected signal is amplified at 4K and at room temperature, I/Q demodulated to baseband, filtered with a 1 MHz low-pass filter, and recorded with a fast data acquisition card.

\section*{Acknowledgement}
We thank Floor van Riggelen and Luc Blom for software support and Raymond Schouten for electronics support. We thank all members of the QuTech spin qubit group for fruitful discussions.
We acknowledge support by the Netherlands Organization of Scientific Research (NWO) Vici program, a supplement for Topconsortia for Knowledge and Innovation (TKI) of the Dutch Ministry of Economic Affairs, the Swiss National Science Foundation (SNF), and the QuantERA ERA-NET Cofund in Quantum Technologies implemented within the European Union's Horizon 2020 Programme, and a Japan Society for the Promotion of Science (JSPS) Postdoctoral Fellowship for Research Abroad.
This research was sponsored by the Army Research Office (ARO) under grant numbers W911NF-17-1-0274 and W911NF-12-1-0607. The views and conclusions contained in this document are those of the authors and should not be interpreted as representing the official policies, either expressed or implied, of the ARO or the US Government. The US Government is authorized to reproduce and distribute reprints for government purposes notwithstanding any copyright notation herein.

\section*{Author contributions}
C.V. and A.M.J.Z. performed the experiment, following preliminary experiments by T.H and T.F., and analyzed the data. C.V., A.M.J.Z., P.T.E., C.J.v.D. and L.M.K.V. discussed the data. U.M. and J.P.D designed and fabricated the samples. C.R. and W.W. grew the heterostructure. C.V., A.M.J.Z. and L.M.K.V. wrote the manuscript with input from all authors. L.M.K.V. initiated and supervised the project.

\section*{Additional information}
\paragraph*{Data availability:} The datasets generated and/or analysed during the current study are available from the corresponding author on request.
\paragraph*{Competing interests:} The authors declare no competing financial or non-financial interests.

\clearpage

\renewcommand{\thefigure}{S\arabic{figure}}
\setcounter{figure}{0}
\section*{Supplementary Information}

\begin{figure*}
\centering
\includegraphics [width=0.95\linewidth] {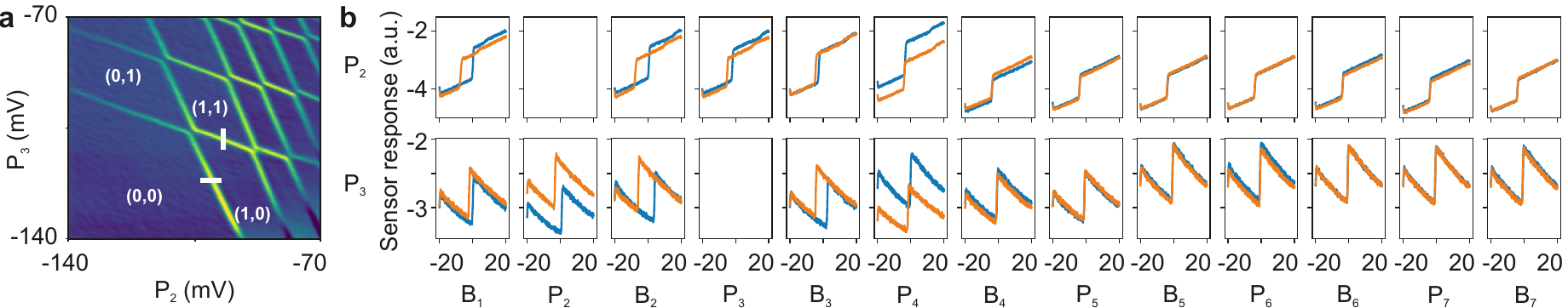}
\caption[]{\textbf{Measuring cross-capacitances.}
(a) Charge stability diagram of a DQD ($\rm{QD}_2$ and $\rm{QD}_3$) in the single electron regime. Cross capacitance has not been corrected, i.e.\ the virtual gates are still equal to the physical gates.
(b) Set of scans to determine the cross-capacitance coupling of each of the gates to the potentials of $\rm{QD}_2$ and $\rm{QD}_3$. The scans of P$_2$ (top row) and P$_3$ (bottom row) are taken along the white lines in (a), for two different values of the gate indicated below each panel. For the orange traces, the corresponding gate voltage has been increased by $\Delta V$ = +10~mV compared to the blue traces. As described in the main text, from this set of measurements the entries of the cross capacitance matrix corresponding to VP$_2$ and VP$_3$ can be determined.
}
\label{figS1}
\end{figure*}

\begin{figure}
\centering
\includegraphics [width=0.9\linewidth] {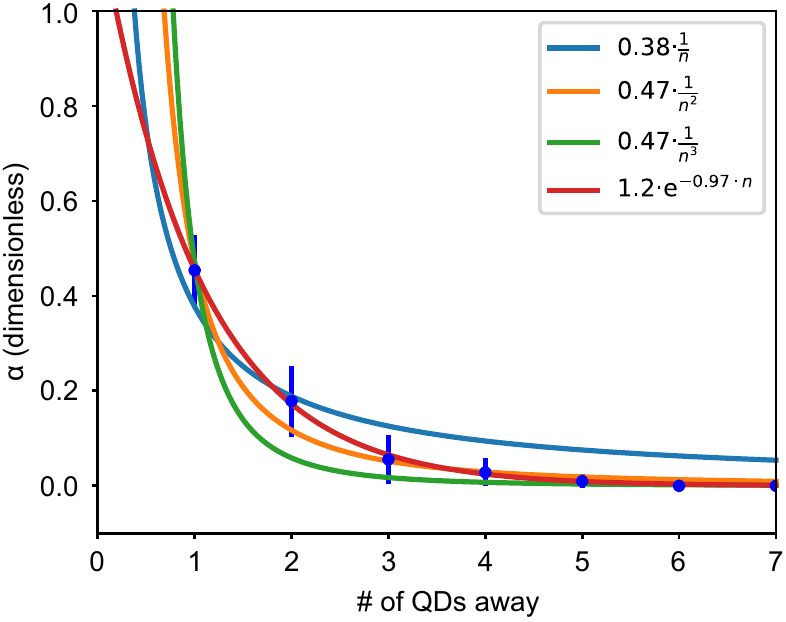}
\caption[]{\textbf{Cross-talk as a function of distance.}
The measured cross-capacitance between plunger gates and QD electrochemical potential as a function of the distance between the gate and the dot. Each data point is the average of the corresponding matrix entries, with the error bar indicating the standard deviation. The data is fitted with several fit functions, as indicated.}
\label{figS2}
\end{figure}

\begin{figure}
\centering
\includegraphics [width=0.95\linewidth] {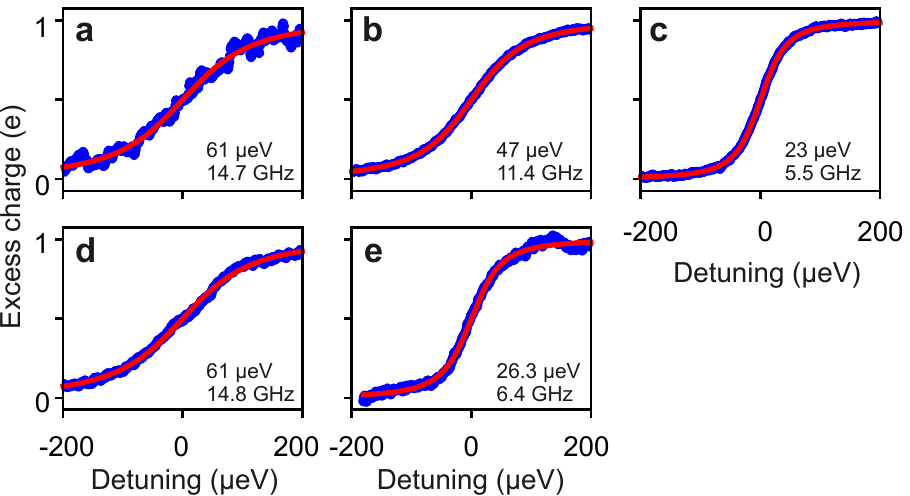}
\caption[]{\textbf{Tunnel coupling.}
Scan along the interdot detuning axis of neighbouring QDs with one electron in these two dots. The data has been fit according to the model described in~\cite{DiCarlo2004,Hensgens2017}. In each panel the  tunnel coupling extracted from the data is given in $\mu$eV and GHz.
}
\label{figS3}
\end{figure}

\begin{figure*}
\centering
\includegraphics [width=0.9\linewidth] {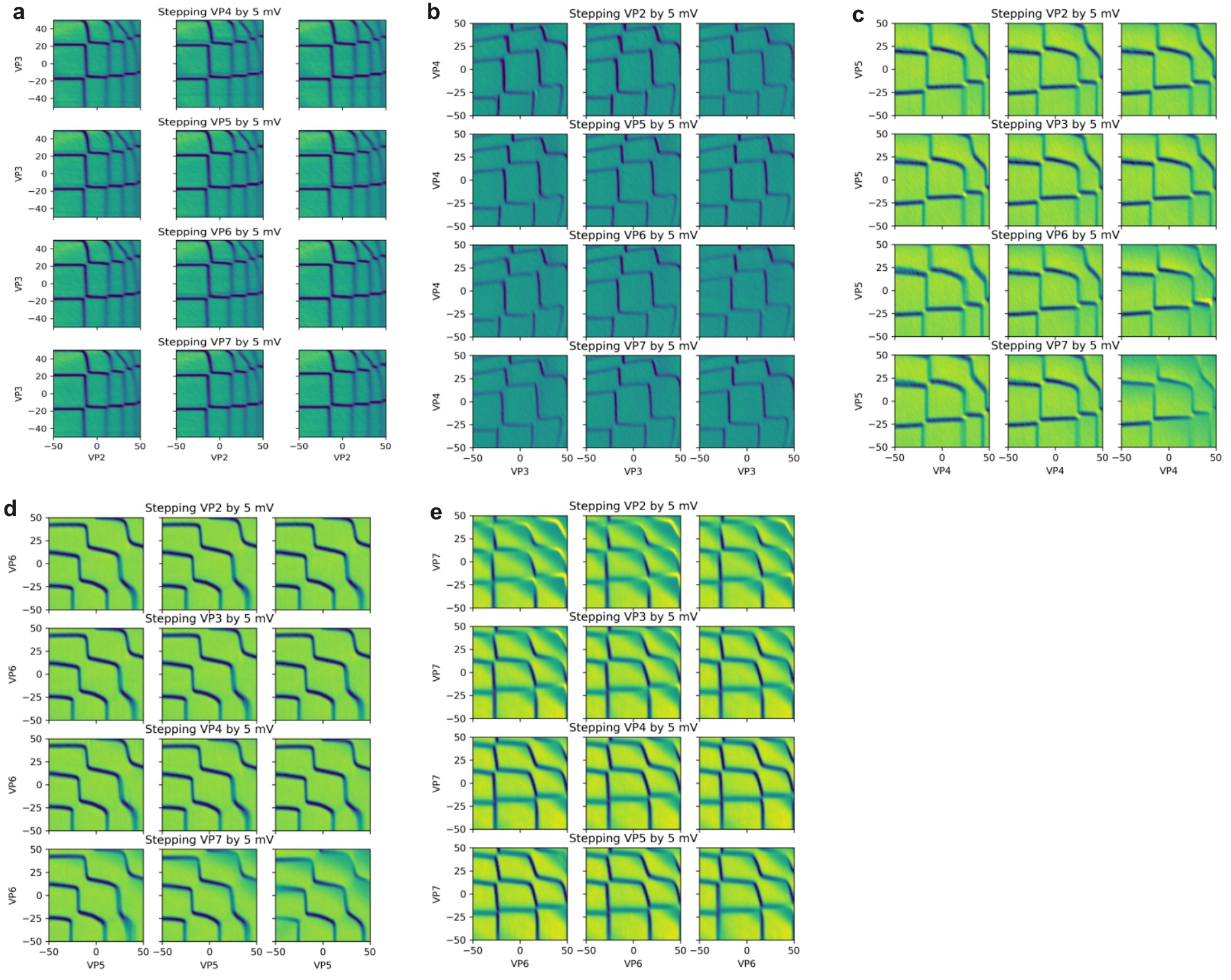}
\caption[]{\textbf{Verification of charge addition lines.}
Charge stability diagrams of (a) $\mathrm{VP}_2$ and $\mathrm{VP}_3$, (b) $\mathrm{VP}_3$ and $\mathrm{VP}_4$, (c) $\mathrm{VP}_4$ and $\mathrm{VP}_5$, (d) $\mathrm{VP}_5$ and $\mathrm{VP}_6$ and (e) $\mathrm{VP}_6$ and $\mathrm{VP}_7$. In each of the panels the other gate voltages have been stepped by $\pm 5$ mV as indicated in the figure.
}
\label{figS4}
\end{figure*}

Figs.~\ref{figS1}-\ref{figS4} and Table~\ref{tab1} show additional data and information, referred to in the main text.

\begin{figure}
\centering
\includegraphics [width=0.9\linewidth] {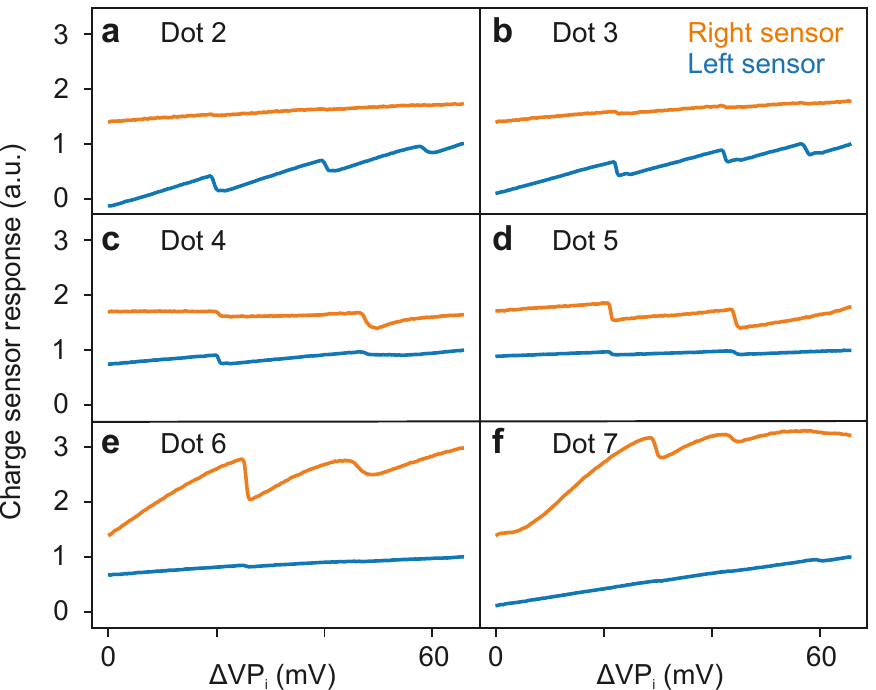}
\caption[]{\textbf{1D traces.}
(a-f) Consecutive fast sweeps of the virtual plunger gates of the sextuple dot. The raw demodulated signal of both sensors is shown.
}
\label{figS5}
\end{figure}

Fig.~\ref{figS5} shows an alternative visualization of the charge transitions of the sextuple QD. Performing just 1D sweeps of the virtual plunger gates, which are a horizontal or vertical cut through a 2D charge stability diagram, allows measuring all charge transitions with a refresh rate of a few Hz. This allows to control parameters of the sextuple dot and verify the effect on all the charge occupations of all six dots in real-time. The data shows that the left charge sensor is most sensitive to charge transitions of QD$_2$ and its sensitivity decreases for dots farther away. It is practically insensitive to charge transitions in QD$_6$ and QD$_7$. Both charge sensors are thus necessary to read out the full device. While the 1D traces can be recorded very fast, they lack certain information and are more prone to incorrect interpretation. For instance, in 1D traces, interdot tunnel couplings cannot be extracted and the difference between interdot and dot-reservoir transitions may or may not be easily seen. 2D charge stability diagrams remove such ambiguities or limitations.

\begin{table*}
\begin{tiny}
\begin{tabular}{|c|c|c|c|c|c|c|c|c|c|c|c|c|c|c|c|c|c|c|c|c|c|c|c|c|}
\hline
Gate names & $D$ & $B_0$ & $P_1$ & $B_1$ & $P_2$ & $B_2$ & $P_3$ & $B_3$ & $P_4$ & $B_4$ & $P_5$ & $B_5$ & $P_6$ & $B_6$ & $P_7$ & $B_7$ & $P_8$ & $B_8$ & $Y_0$ & $X_1$ & $Y_1$ & $Y_2$ & $X_2$ & $Y_3$\\ \hline
Bias cooling & +100 & +50 & +50 & +100 & +100 & +100 & +100 & +200 & +100 & +250 & +100 & +200 & +100 & +250 & +100 & +200 & +150 & +200 & +50 & +200 & +100 & +50 & +200 & +200 \\ \hline
8dot configuration & -340 & -200 & -295 & -120 & -200 & -25 & -385 & -40 & -240 & -85 & -220 & -260 & +10 & -105 & -75 & -265 & 50 & -230 & -420 & 115 & -515 & -370 & 0 & -500 \\
\hline
\end{tabular}
\end{tiny}
\caption[]{\textbf{} Voltages in mV applied to the different gates during cooldown and typical gate voltages applied in the Qubyte configuration (all eight QDs tuned). The voltages applied during cooldown were chosen based on measurements from an earlier cooldown of the same device.}
\label{tab1}
\end{table*}

%\clearpage

\end{document}